# Guidelines for reporting the use of gel electrophoresis in proteomics

Frank Gibson[1], Leigh Anderson[2], Gyorgy Babnigg[3], Mark Baker[4], Matthias Berth[5], Pierre-Alain Binz[6,7], Andy Borthwick[8], Phil Cash[9], Billy W Day[10], David B Friedman[11], Donita Garland[12], Howard B Gutstein[13], Christine Hoogland[6], Neil A Jones[14], Alamgir Khan[4], Joachim Klose[15], Angus I Lamond[16], Peter F Lemkin[17], Kathryn S Lilley[18], Jonathan Minden[19], Nicholas J Morris[1], Norman W Paton[20], Michael R Pisano[21], John E Prime[22], Thierry Rabilloud[23], David A Stead[24], Chris F Taylor[25,26], Hans Voshol[27], Anil Wipat[28] & Andrew R Jones[29]

1. Institute for Cell and Molecular Biosciences, The Medical School, University of Newcastle, Newcastle upon Tyne, UK.
2. Plasma Proteome Institute, PO Box 21466, Washington, DC 20009-1466, USA.
3. Argonne National Laboratory, 9700 S. Cass Ave., Argonne, Illinois 60439, USA.
4. Australian Proteome Analysis Facility Ltd and Department of Chemistry & Biomolecular Sciences, Macquarie University, Sydney, NSW 2109, Australia.
5. Decodon, GmbH W.-Rathenau-Str., 49a, 17489 Greifswald, Germany.
6. Swiss Institute of Bioinformatics, 1 Rue Michel-Servet, CH-1211 Geneva 4, Switzerland.
7. GeneBio SA, 25 Av. de Champel, CH-1206 Geneva, Switzerland.
8. Nonlinear Dynamics, Cuthbert House, All Saints, Newcastle-upon-Tyne, NE1 2ET Newcastle, UK.
9. Department of Medical Microbiology, University of Aberdeen, Foresterhill, Aberdeen AB25 2ZD, UK.
10. Department of Pharmaceutical Sciences, Department of Chemistry, Proteomics Core Lab University of Pittsburgh, Pittsburgh, Pennsylvania 15213, USA.
11. Mass Spectrometry Research Center, Proteomics Laboratory, 465 21st Ave S. Room 9160, Medical Research Building III, Vanderbilt University, Nashville, Tennessee 37232, USA.
12. National Eye Institute, National Institutes of Health, 9000 Rockville Pike, Bethesda, Maryland 20892, USA.
13. Departments of Anesthesiology and Molecular Genetics, University of Texas-MD Anderson Cancer Center, 1515 Holcombe Blvd., Houston, Texas 77030, USA.
14. Disease & Biomarker Proteomics, Genomic and Proteomic Sciences, Genetics Research, GlaxoSmithKline R&D, Stevenage, Herts SG1 2NY, UK.

15. Charité-Universitaetsmedizin Berlin, Institute of Human Genetics, D-13353 Berlin, Germany.

16. Wellcome Trust Biocentre MSI/WTB Complex, University of Dundee, Dow Street, Dundee, DD1 5EH, UK.

17. National Cancer Institute, Building 469, Room 150B, Frederick, Maryland 21702, USA.

18. Cambridge Centre for Proteomics, Department of Biochemistry, University of Cambridge, Tennis Court Road, Cambridge, CB2 1QR, UK.

19. Carnegie Mellon University, 4400 Fifth Avenue, Pittsburgh, Pennsylvania 15213, USA.

20. School of Computer Science, University of Manchester, Oxford Road, Manchester M13 9PL, UK.

21. Proteomic Research Services, Inc., 4401 Varsity Drive, Suite E, Ann Arbor, Michigan 48108, USA.

22. KuDOS Pharmaceuticals, 327 Cambridge Science Park, Milton Road, Cambridge, CB4 0WG, UK.

23. DRDC/ICH, INSERM U548, CEA-Grenoble, 17, rue des martyrs, F-38054 Grenoble, CEDEX 9, France.

24. Aberdeen Proteomics, School of Medical Sciences, University of Aberdeen, Aberdeen, IMS Building, Foresterhill, Aberdeen AB25 2ZD, UK.

25. EMBL Outstation, European Bioinformatics Institute, Wellcome Trust Genome Campus, Hinxton, Cambridge, UK.

26. NERC Environmental Bioinformatics Centre, Mansfield Road, Oxford, OX1 3SR, UK.

27. Novartis Institutes for Biomedical Research, 250 Mass Ave., Cambridge, Massachusetts 02139, USA.

28. School of Computing Science, 8th Floor, Claremont Tower, Newcastle University, Newcastle upon Tyne, NE1 7RU, UK.

29. Department of Pre-clinical Veterinary Science, Faculty of Veterinary Science, University of Liverpool, Liverpool, L69 7ZJ, UK.

Correspondence to: Frank Gibson1 e-mail: frank.gibson@ncl.ac.uk

We wish to alert your readers to the MIAPE Gel Electrophoresis (MIAPE-GE) guidelines specifying the minimum information that should be provided when reporting the use of n-dimensional gel electrophoresis in a proteomics experiment. Developed through a joint effort between the gel-based analysis working group of the Human Proteome Organisation's Proteomics Standards Initiative (HUPO-PSI; http://www.psidev.info/) and the wider proteomics community, they constitute one part of the overall Minimum Information about a Proteomics Experiment (MIAPE) documentation system published last August in Nature Biotechnology [1].

MIAPE-GE comprises a checklist of information that should be provided about gel electrophoresis performed in the course of generating a data set that is submitted to a public repository or when such an experimental step is reported in a scientific publication (for instance, in the materials and methods section; see Box 1). MIAPE-GE specifies neither the format in which information should be transferred nor the structure of any repository or document. However, HUPO-PSI is not developing the MIAPE modules in isolation; several compatible data exchange standards are now well established and supported both by public databases and by data processing software in proteomics. MIAPE-GE will be implemented by public repositories, such as PRIDE, Swiss2DPage and Gelbank, and the PSI's GelML data format is designed to support MIAPE-GE-compliant submission [2].

Gel electrophoresis facilitates the separation of protein (or peptide) mixtures, usually in a gel matrix under the application of an electric field. MIAPE-GE contains a glossary (Supplementary Table 1 online) specifying the minimum information to report about a gel electrophoresis experiment so as to enable the extraction of the maximum value from data generated, specifically addressing: gel matrix manufacture and preparation; running conditions; visualization techniques, such as staining; the method of image capture; and a technical description of the image obtained. The module does not explicitly cover sample preparation, although it requires the recording of which samples were loaded onto a gel and whether the protein complement had been labeled. Neither does the module cover the informatics process or the analysis of digitized gel images; this is addressed in a separate module, MIAPE-GI (Gel Informatics). These and other items falling outside the scope of this module may be captured in complementary modules, the latest versions of which can be obtained from the MIAPE home page.

These guidelines are intended to evolve, and readers are directed to MIAPE homepage (http://www.psidev.info/miape/) to check compliance with the most up-to-date version. They may also view the most recent version of MIAPE-GE at the module's homepage (http://www.psidev.info/miape/ge/)

References

1. Taylor, C.F. et al. Nat. Biotechnol. 25, 887–893 (2007).
2. Jones, A.R. & Gibson, F. Proteomics 7 Suppl 1, 35–40 (2007).

Box 1 Content snapshot for MIAPE-GE

The full MIAPE-GE document is divided into three parts: an introduction providing background and context; a summary list of the items to be reported; and a glossary with definitions and examples.

The MIAPE-GE guidelines themselves are subdivided as follows:

1. General features. Initiation date; contact information for the data set; type of electrophoresis.

2. Sample. The material applied to the gel matrix and its role; labels or tags used; loading buffer.

3. Gel matrix and electrophoresis. Physicochemical components and properties of the gel matrix; electrophoresis protocol.

4. Inter-dimension process. Any process or processes carried out between the running of separation dimensions, such as equilibration, or reduction and alkylation.

5. Detection process. Examples include direct methods such as staining proteins on the gel and indirect methods such as exposing a gel matrix containing a radiolabeled sample to photographic film or the transfer of proteins to an alternate matrix (e.g., immunoblotting).

6. Image acquisition. Equipment and procedure used to capture a digitized representation of an electrophoresed gel matrix and sample, or a detection medium.

7. Image. Descriptors for the digitized image produced as a result of the Image Acquisition, such as name and dimensions, resolution and bit-depth.